\documentclass[twocolumn,preprintnumbers,amsmath,amssymb]{revtex4}
\usepackage{graphicx}% Include figure files
\usepackage{dcolumn}% Align table columns on decimal point
\usepackage{bm}% bold math

\newcommand{\nc}{\newcommand}
\nc{\beq}{\begin{equation}}
\nc{\eeq}{\end{equation}}
\nc{\barray}{\begin{eqnarray}}
\nc{\earray}{\end{eqnarray}}
\nc{\barrayn}{\begin{eqnarray*}}
\nc{\earrayn}{\end{eqnarray*}}
\nc{\bcenter}{\begin{center}}
\nc{\ecenter}{\end{center}}
\nc{\ket}[1]{| #1 \rangle}
\nc{\bra}[1]{\langle #1 |}
\nc{\0}{\ket{0}}
\nc{\mc}{\mathcal}
\nc{\mcp}{\mc{P}}
\nc{\mcm}{\mc{M}}
\nc{\mcd}{\mc{D}}
\nc{\gev}{\;\mathrm{GeV}}
\nc{\er}[1]{(\ref{eq:#1})}
\nc{\onehalf}{\frac{1}{2}}
\nc{\partialbar}{\bar{\partial}}
\nc{\psit}{\widetilde{\psi}}
\nc{\Tr}{\mbox{Tr}}
\nc{\spacer}{\phantom{spacer}}

\def\chii0{\chi_i^0}
\def\chij0{\chi_j^0}

\def\st{\tilde{t}}
\nc{\infinity}{\infty}
\nc{\ttbar}{t\bar t}

\begin{document}

\preprint{RUNHETC-2008-21}

\title{Polarized tops from new physics: signals and observables}

\author{Jessie Shelton}
\email{jshelton@physics.rutgers.edu}
\affiliation{%
NHETC, Rutgers Department of Physics and Astronomy\\
}%

\begin{abstract}

Top quarks may be produced in large numbers in association with new physics at the LHC.
The polarization of these top quarks probes the chiral structure of the new physics.
We discuss several kinematic distributions which are sensitive to the polarization of 
single top quarks and can be used without full event reconstruction.  For collimated tops 
we construct polarization-sensitive observables for both hadronic and leptonic decay modes
and plot their distributions.  We compute the observable polarization signals from 
top quarks produced in the on-shell cascade decay of a stop squark into a
top quark and a neutralino, as well as top quarks produced in the analogous decay chain in 
same-spin partner models.

\end{abstract}

\maketitle

%%%%%%%%%%%%%%%%%%%%%%%%%%%%%%%%%%%%%%%%%%%%%%%%%%%%%%%%%%%%%%%%%%%%%%%%%%
\section{Introduction}
%%%%%%%%%%%%%%%%%%%%%%%%%%%%%%%%%%%%%%%%%%%%%%%%%%%%%%%%%%%%%%%%%%%%%%%%%%

The LHC will provide a wealth of data on the physics of the top quark
\cite{tdrs,top@LHC}.
Standard Model pair production $pp\to \ttbar$ alone is approximately
830 picobarns, allowing for detailed study of top
properties, such as spin correlations and rare decays.  In addition,
any new physics which is responsible for electroweak symmetry breaking
must couple strongly to the top quark, leading to many events where
top quarks are produced in association with new physics.

Models which aim to stabilize the electroweak hierarchy typically
predict a top quark partner, such as the stop squark in SUSY or the
$T'$ in little Higgs models.  In order to suppress contributions to
precision electroweak observables, this top partner is frequently
taken to be odd under a discrete parity, which in turn typically leads
to a new stable invisible particle.  Thus a standard signature of a
broad class of well-motivated models is the production of a single top
quark in the cascade decay of an on-shell top partner, accompanied by
additional particles in the event and possibly missing energy.  

The coupling of top partners to top quarks is in general chiral, and a top 
quark produced in the cascade decay of a top partner will be polarized to
a degree which depends on the kinematics of the decay chain as well as
the couplings.  The top undergoes weak decay before the process of
hadronization can dilute information about its polarization, so the
angular distributions of its daughter particles reflect the
polarization of the parent top.  Observables which measure single top
polarization can therefore yield measurements which are sensitive to
the chiral structure of any new physics sector.

We focus here on kinematic distributions which are can be used without fully
reconstructed events.  To fully reconstruct events with missing energy 
one must make assumptions about the event topology and the identities of 
particles not directly observed.  This is not always desirable in events 
with high multiplicity final states and large missing energy.  In such events
hadronic top decays will likely be more useful than leptonic top
decays for both event selection and top reconstruction, especially in the
initial stages of analysis.  Fully collimated tops allow polarization
study without full event reconstruction in {\em both} leptonic and hadronic
decay modes, as in the collinear limit one may construct observables which become
independent of the unknown boost between the top rest frame and the lab frame.

In section~\ref{sec:obs} we discuss several observables whose distributions 
can measure single top polarization, for both collimated and uncollimated top quarks.
In section~\ref{sec:signal} we compute the polarization of top quarks produced
from the cascade decay of a top partner to a top quark and an additional particle
which is not directly observed.  In section~\ref{sec:assembly}, we relate this 
polarization to experimentally observable signals, and in 
section~\ref{sec:theend} we conclude.

%%%%%%%%%%%%%%%%%%%%%%%%%%%%%%%%%%%%%%%%%%%%%%%%%%%%%%%%%%%%%%%%%%%%%%%%%%
\section{Kinematic distributions from Standard Model top decay}
%%%%%%%%%%%%%%%%%%%%%%%%%%%%%%%%%%%%%%%%%%%%%%%%%%%%%%%%%%%%%%%%%%%%%%%%%%
\label{sec:obs}

In the Standard Model the top decays almost entirely through $t\to
Wb\to f_1 \bar f_2 b$.  The angular distributions of the final state
particles depend on the spin of the top.  In general the angular
distribution of any one of the top's three daughter particles can be written
\begin{equation}
\label{eq:top-general}
\frac{1}{\Gamma} \frac{d\Gamma}{d \cos \theta_f} = \onehalf \left( 1+\mcp_t 
           \kappa_f \cos \theta_f \right),
\end{equation}
where $\theta_f$ is the angle between the daughter fermion $f $ and
the top spin axis, as measured in the top rest frame, and $\mcp_t$ is
the polarization of the top along that axis. We use conventions where
a top with spin $\pm 1/2 $ has $\mcp_t= \pm 1$.  The ``spin analyzing
power'' of the fermion $f$, $\kappa_f $, is a calculable coefficient,
which depends on the identity of $f $.  Unitarity restricts $|\kappa_f
| \leq 1 $.

The anti-fermion from the $W$ decay---the charged lepton, in leptonic
$W$ decays---is maximally correlated with the top spin, $\kappa_{\bar
f_2} = 1$ \cite{Jezabek:1988ja}.  Therefore the charged lepton is the
natural first object to consider in a study of top polarization.
However, in many inclusive final states, $pp\to t X$, the leptonic
decay mode may be of limited utility, as the neutrino makes top
reconstruction indirect.  While it is possible to study top polarization 
without reconstructing the top rest frame, for instance with highly 
boosted tops, the most direct window into top polarization uses 
knowledge of the top rest frame.  Moreover,
depending on the particular study, top identification may proceed only
via hadronic decays where three jets can be combined within the
appropriate invariant mass window.  It is therefore interesting to consider 
polarization-sensitive observables that can be constructed for 
hadronic top quarks.

Hadronically decaying tops present three natural candidate objects
whose angular distributions measure the polarization of the parent top.  
One is the $b$-jet.  At
tree level, the distribution of the $b$-quark is given by equation
\er{top-general} with
\begin{equation}
\kappa_b = - \frac{m_t ^ 2-2 m_W ^ 2}{m_t ^ 2+2 m_W ^ 2} \simeq -0.4.
\end{equation}
One can also consider the angular distribution of the (reconstructed)
$W$ boson, which is of the form~(\ref{eq:top-general}) with
\beq
\kappa_W = -\kappa_b\simeq 0.4.
\eeq
The spin analyzing power of the $b$-jet (or the reconstructed $W$) is 
less than half of the spin analyzing power of the anti-fermion from 
the decay of the $W$.  Unfortunately, in hadronic decays, the down-quark jet cannot be
distinguished from the up-quark jet.  However, in the top rest frame
the down quark is on average less energetic than the up quark.  
Thus the less energetic of the two light quark jets in the top rest frame 
constitutes another interesting object whose angular distribution can serve 
to measure the top quark polarization.  At tree level, this jet has a net 
spin analyzing power \cite{Jezabek:1994qs,qcd-corr-had}
\begin{equation}
\kappa_{j} \simeq 0.5,
\end{equation}
where, again, $j $ denotes the less energetic of the two light quark jets 
in the top rest frame.  Leading QCD corrections to $\kappa_b $ and $\kappa_j $ 
have been calculated, and are of order a few percent \cite{qcd-corr-had}.  In
both cases the effect of the QCD corrections is to decrease $|\kappa|$.
 
Depending on the method of top identification used in a given study, the angular distributions of
one or several of the three objects mentioned here (the $b$-jet, the reconstructed 
$W $, or the softer of the two light 
quark jets) can form attractive observables which are sensitive to the top quark
polarization along the top quark direction of motion in the lab frame.  

%%%%%%%%%%%%%%%%%%%%%%%%%%%%%%%%%%%%%%%%%%%%%%%%%%%%%%%%%%%%%%%%%%%%%%%%%%
\subsection{Collimated top quarks}
%%%%%%%%%%%%%%%%%%%%%%%%%%%%%%%%%%%%%%%%%%%%%%%%%%%%%%%%%%%%%%%%%%%%%%%%%%

Suitably massive new physics may produce daughter top quarks which are highly 
boosted in the lab frame.  The top decay products will then be partially 
or fully collimated, necessitating novel approaches to top identification and
reconstruction \cite{boostedtops0,boostedtops,Almeida:2008tp}.  When the tops are 
highly collimated, the 
angular distributions discussed above cannot be well measured
and other kinematic variables are more appropriate.

\begin{figure} 
\begin{center} 
\includegraphics[width=8.0cm]{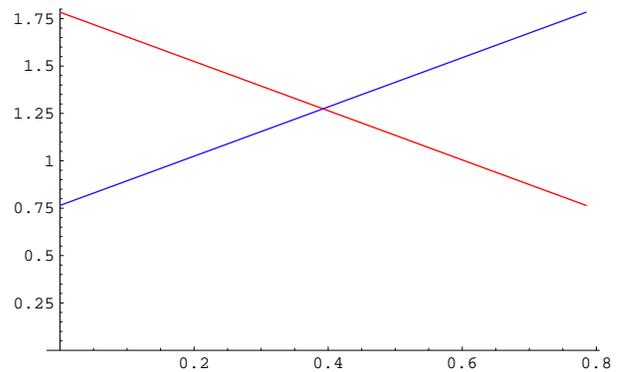} 
\end{center} 
%\vspace{-0.75cm} 
\begin{small}
\caption{The distribution $1/\Gamma \, d\Gamma/dz$ of the fraction of visible 
lab frame energy carried by the $b$-jet in a highly boosted hadronic top,
$z = \mc{E}_b/\mc{E}_t $, plotted as a function of $z $.  The blue line (weighted towards high values of $z$)
is the distribution for a negative-helicity top quark; the red line (weighted towards
low values of $z $) is the distribution for a positive-helicity top quark.
\label{fig:bdist}}
\end{small}
\end{figure}

We first consider collimated hadronic tops.  When the $b$-jet (or equivalently the $W$) 
can be separately identified within the top, then the natural 
polarization-sensitive observable is the fraction of the
lab-frame top energy carried by the $b$-jet,
\beq
z = \frac{\mc{E}_b }{\mc{E}_t},
\eeq
where $\mc{E}_f$ denotes the energy of a particle $f$ in the lab frame.
Taking the axis of polarization to be given by the top direction of motion, 
the variable $z$ is simply related to $\cos\theta_b$,
\beq
\cos\theta_b =\frac{1}{\beta}\left(\frac{2m_t^ 2}{m_t^2-m_W ^ 2} z-1\right),
\eeq
where $\beta$ is the boost between the top rest frame in the lab frame (we work 
to tree level and in the narrow width approximation).  The distribution of $z$ 
is given by
\beq
\frac{1}{\Gamma} \frac{d\Gamma}{d z} = \frac{m_t^ 2}{m_t^2-m_W ^ 2}(1-\kappa_b \mcp_t ) 
     + \kappa_b \mcp_t \frac{2m_t^ 2}{m_t^2-m_W ^ 2} z
\eeq
in the collinear limit $\beta \to 1$.
This distribution is plotted in Figure~\ref{fig:bdist} for $\mcp_t = \pm 1$.
Exactly analogous distributions pertain for the reconstructed $W $,
interchanging the roles of positive and negative helicities.
The distribution of the total $p_T$ of the $b$-jet has also been proposed as a 
polarimeter for highly boosted top quarks \cite{Almeida:2008tp}.

%%%%%%%%%%%%%%%%%%%%%%%%%%%%%%%%%%%%%%%%%%%%%%%%%%%%%%%%
%% Figure: distribution of b energy fraction for collimated hadronic tops
%%%%%%%%%%%%%%%%%%%%%%%%%%%%%%%%%%%%%%%%%%%%%%%%%%%%%%%%
%%%%%%%%%%%%%%%%%%%%%%%%%%%%%%%%%%%%%%%%%%%%%%%%%%%%%%%%

\begin{figure} 
\begin{center} 
\includegraphics[width=8.0cm]{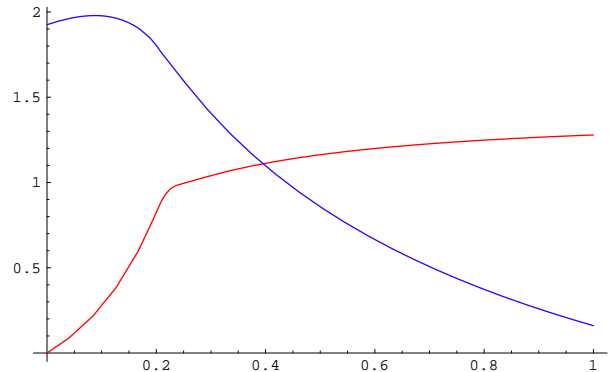} 
\end{center} 
%\vspace{-0.75cm} 
\begin{small}
\caption{The distribution $1/\Gamma \, d\Gamma/du$ of the fraction of visible 
lab frame energy carried by the lepton in a highly boosted leptonic top,
$u = \mc{E}_\ell/(\mc{E}_\ell+\mc{E}_b) $, plotted as a function of $u $.  The blue curve (peaked towards low values of $u$)
is the distribution for a negative-helicity top quark; the red curve (weighted towards
high values of $u $) is the distribution for a positive-helicity top quark.
\label{fig:udist}}
\end{small}
\end{figure}

In the highly relativistic limit, leptonic tops can be used to study polarization without
any need for additional event reconstruction, as observables can be constructed which 
in the collinear limit are independent of the unknown boost between the top rest frame and 
the lab frame.  Collimated leptonic top quarks allow the energy of the lepton and the $b$-jet to be separately 
measured.  The momentum fraction carried by the lepton provides a natural polarimeter.  
Consider the observable
\beq
u =\frac{\mc{E}_\ell}{\mc{E}_\ell+\mc{E}_b},
\eeq
the fraction of the visible lab-frame energy carried by the lepton.  This can be 
expressed in terms of top rest-frame quantities as
\begin{widetext}
\beq
u =\frac{2 m_t E_\ell(1+\beta \cos\theta_\ell)}
    {2 m_t E_\ell(1+\beta \cos\theta_\ell) + (m_t ^ 2-m_W ^ 2)
    (1+\beta\cos\theta_\ell\cos\zeta +\beta\sin\theta_\ell\sin\zeta\cos\alpha)},
\eeq
\end{widetext}
where $\theta_\ell$ is the angle between the lepton momentum and the top direction 
of motion, $E_\ell$ is the lepton energy in the top rest frame, $\alpha$ describes 
the orientation of the $b$-$\ell$ plane relative to the top direction of motion, and 
\beq
\cos\zeta = \frac{m_t m_W ^ 2-E_\ell (m_t ^ 2+m_W ^ 2)}{ E_\ell(m_t ^ 2-m_W ^ 2)}
\eeq
is the angle between the lepton and the $b$ momenta in the $b$-$\ell$ plane. The (unknown)
boost from the top rest frame to the lab frame is parameterized by $\beta$, which we
will henceforth take to unity.  In these coordinates, the matrix element for
top decay is
\beq
|\mc{M} |^ 2\propto (m_t E_\ell-2 E_{\ell} ^ 2) (1+\mcp_t \cos \theta_\ell),
\eeq
where again we take the axis of spin quantization to lie along the direction of top motion.
The distribution of $u $ can then be obtained from
\barray
\nonumber
\frac{1}{\Gamma} \frac{d\Gamma}{d u} &\propto & \int dE_\ell \,d\cos\theta_\ell \, d\alpha \, |\mc{M} |^ 2
    (\mcp_t, E_\ell,\cos\theta_\ell) \times \\
       &&\spacer \times \delta (u-u (E_\ell,\cos\theta_\ell,\alpha)).
\earray
We use the delta function to carry out the integral over $\alpha$, and evaluate the
remaining integrals numerically.  The resulting distributions for positive- and 
negative-helicity top quarks are plotted in Figure~\ref{fig:udist}.  The distributions 
have a shoulder at 
\beq
u = \frac{2 m_t E_{\ell,min}}{2 m_t E_{\ell,min}+ (m_t ^ 2-m_W ^ 2)} =\frac{m ^ 2_W}{m_t ^ 2} \simeq 0.215,
\eeq
where $E_{\ell,min}=m^2_W/(2 m_t)$ is the minimum possible value of the lepton energy in the
top rest frame.  Above this value of $u $, the integrand can receive contributions from multiple
angular configurations as $E_\ell$ varies, while below this value of $u $, only limited angular 
configurations can contribute.

%%%%%%%%%%%%%%%%%%%%%%%%%%%%%%%%%%%%%%%%%%%%%%%%%%%%%%%%
%% Figure: distribution of lepton energy fraction for collimated leptonic tops
%%%%%%%%%%%%%%%%%%%%%%%%%%%%%%%%%%%%%%%%%%%%%%%%%%%%%%%%
%%%%%%%%%%%%%%%%%%%%%%%%%%%%%%%%%%%%%%%%%%%%%%%%%%%%%%%%

%%%%%%%%%%%%%%%%%%%%%%%%%%%%%%%%%%%%%%%%%%%%%%%%%%%%%%%%%%%%%%%%%%%%%%%%%%
\section{Single top polarization from cascade decays}
%%%%%%%%%%%%%%%%%%%%%%%%%%%%%%%%%%%%%%%%%%%%%%%%%%%%%%%%%%%%%%%%%%%%%%%%%%
\label{sec:signal}

Cascade decays of top partners to tops + missing energy are a
prototypical example of the processes for which the above
observables are well-suited.  In this section we compute the polarization 
of top quarks produced
from the decay of a top partner into a top quark and a boson partner, for both 
SUSY models (where the boson partner is a neutralino) and same-spin
partner models (where the boson partner is a massive vector boson).  We work 
to tree level and in the narrow width approximation, and take the top quark to 
have standard model decays.  The polarization $\langle\mcp_P\rangle $ 
is computed along the
axis determined by the direction of the top motion in the parent rest
frame, which is not directly observable; in section~\ref{sec:assembly}
we will show how to relate this to the polarization $\langle\mcp_D \rangle$ 
along the axis determined by the direction of the top motion in the laboratory 
frame, which yields predictions for experimentally observable distributions.

\subsection{Production polarization in SUSY decay chains}

Consider the two-body decay of an on-shell stop squark $\st $ into a
top quark and a neutralino $\chi ^ 0$.  The coupling for this process
can be parameterized as
\[
-\mc{L}_{int} = a\, \st \,t_L\chi^0+b\,\st\, \bar t_R\bar \chi^ 0 +\mathrm{H.c.},
\]
where the Yukawa couplings $a $ and $b $ are in general not equal.
The top quark is thus produced with some nonzero polarization,
the magnitude of which depends on (1) the masses of the particles in the 
decay chain and
(2) the mixing of the stop squarks and the neutralinos.  We take the
axis of spin quantization to lie among the top direction of motion in
the parent stop rest frame.

The production amplitudes for positive and negative helicity top quarks 
depend on the following functions of the top
and neutralino energy-momenta,
\[
F_\pm = \frac{(E_t+m_t\pm |p_t|) (E_\chi +m_\chi\pm |p_\chi|)}
                {\sqrt{4(E_t+m_t) (E_\chi +m_\chi)}},
\]
where all quantities are to be evaluated in the stop rest frame.
For this on-shell process, the top and neutralino energy-momenta are 
fixed in terms of the masses of the particles in the cascade.
The functions $F_\pm$ result simply from explicit evaluation of the usual
spinor wavefunction operators $\sqrt{p\cdot \sigma} $, $\sqrt{p\cdot
\bar \sigma} $ on helicity eigenspinors.  They measure the separate
contributions of the two helicity states in the left- or right-handed
fermion wave function.  The matrix element to produce a positive
(negative)-helicity top quark is then the sum of the contributions
from the left- and right-handed tops,
\[
\mc M_\pm =\mp i (a F_\mp+b F_\pm).
\]
For finite $m_t$ this gives, as expected, a nonvanishing amplitude for
top quarks of both helicities even in the limit of a purely chiral
vertex, $a\to 0$ or $b\to 0$.  We find the net polarization along the
production axis to be the
\begin{equation}
\langle\mcp_P\rangle =  \frac{ ( | b | ^ 2- | a| ^ 2) M |p_t|}
            {( | a| ^ 2+ | b | ^ 2)(M E_t-m_t ^ 2) +2\mathrm{Re} (a\bar b) m_\chi m_t},
\end{equation}
where $M $ is the mass of the initial squark.  The net polarization $\langle\mcp_P\rangle $
arising from this cascade decay is plotted in the solid curves in
Figure~\ref{fig:avgp} as a function of the masses of the stop squark
and the neutralino.  For simplicity we take purely chiral couplings
($b=1,\; a=0$) in the figure, so the plotted polarization is an upper
bound.

%%%%%%%%%%%%%%%%%%%%%%%%%%%%%%%%%%%%%%%%%%%%%%%%%%%%%%%%%%%%%%%%%%%%%%%%%%
\subsection{Production polarization in same-spin partner model decay chains}
%%%%%%%%%%%%%%%%%%%%%%%%%%%%%%%%%%%%%%%%%%%%%%%%%%%%%%%%%%%%%%%%%%%%%%%%%%

\begin{figure} 
\begin{center} 
\includegraphics[width=8.0cm]{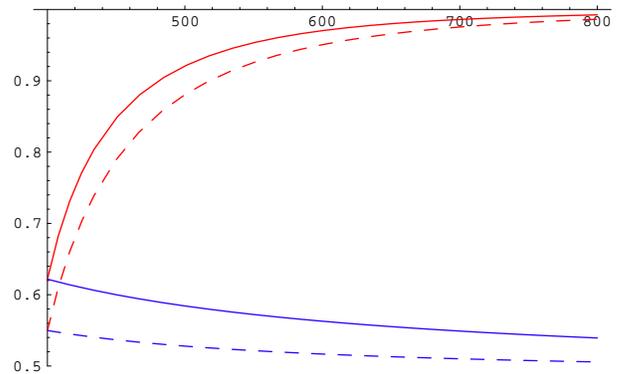} 
\end{center} 
%\vspace{-0.75cm} 
\begin{small}
\caption{Average top quark polarization $\langle \mcp_P\rangle $ as a
function of parent particle mass $M$ in GeV, plotted for purely chiral
couplings ($b = 1,\, a = 0$).  The red (upper) curves have a fixed boson 
partner mass of $200\gev $.  The blue (lower) curves have
a boson partner mass of $M-200\gev$.  Solid (dashed) lines
plot the average polarization for opposite- (same-) spin partners.
\label{fig:avgp}}
\end{small}
\end{figure}

We now consider the analogous decay chain in the case where the top
partner $\hat T $ has spin 1/2 and the vector boson partner $\hat
V_\mu $ has spin 1.  We parameterize this process by
\[
-\mc{L}_{int} = (a \,\bar t\gamma ^\mu\mc{P}_L\hat T+b\,\bar t\gamma ^\mu\mc{P}_R\hat T)\hat V_\mu +\mathrm{H.c.}.
\]
Again, the couplings $a $ and $b $ are in general unequal and depend
on the mixings of the top and vector boson partners.  The computation
is analogous to the one in the previous subsection, except now we must
also average over the spins of the top partner $\hat T$ and sum over
the spins of the vector boson partner $\hat V$.  We find the net
degree of polarization along the production axis to be
\begin{equation}
\langle\mcp_P\rangle =  \frac{ ( | b | ^ 2- | a| ^ 2) p_t (M ^ 2 +2 m_V ^ 2-m_t^ 2)}
            {( | a| ^ 2+ | b | ^ 2)(3 E_tm_V ^ 2+2 Mp_t ^ 2)- 6\mathrm{Re} (a\bar b) m_V ^ 2 m_t},
\end{equation}
where all kinematic quantities ($E_t, p_t$) are to be evaluated in the
parent $\hat T$ rest frame.  A similar computation 
was carried out in \cite{Nojiri:2008ir} for a benchmark model point.
The net polarization
$\langle\mcp_P\rangle $ arising from this cascade decay is plotted in
the dashed curves in figure~\ref{fig:avgp} as a function of the masses
of the top partner and the vector boson partner.  The kinematic
suppression of the ``wrong'' polarization state is less for the
fermion-fermion-vector interaction than it is for the
fermion-fermion-scalar interaction.  That is, for a given set of masses
and couplings, the top polarization predicted by a same-spin cascade
is less than that predicted by an opposite-spin cascade.  This is due
to the additional suppression of the ``wrong'' helicity contribution
which comes from the spinor wave function of the neutralino in the
opposite-spin cascades; in the same-spin case, there is some kinematic
suppression of the ``wrong'' helicity contributions coming from the
longitudinal polarizations of the vector boson partner, but no
suppression from the transverse polarizations.

%%%%%%%%%%%%%%%%%%%%%%%%%%%%%%%%%%%%%%%%%%%%%%%%%%%%%%%%
%% Figure: net polarization as a function of m_T (pure chiral)
%%%%%%%%%%%%%%%%%%%%%%%%%%%%%%%%%%%%%%%%%%%%%%%%%%%%%%%%
%%%%%%%%%%%%%%%%%%%%%%%%%%%%%%%%%%%%%%%%%%%%%%%%%%%%%%%%

%%%%%%%%%%%%%%%%%%%%%%%%%%%%%%%%%%%%%%%%%%%%%%%%%%%%%%%%%%%%%%%%%%%%%%%%%%
\section{Observable top polarization signals}
%%%%%%%%%%%%%%%%%%%%%%%%%%%%%%%%%%%%%%%%%%%%%%%%%%%%%%%%%%%%%%%%%%%%%%%%%%
\label{sec:assembly}

In events where the top is reconstructed but the rest of the event is
not, the natural axis for the top polarization is the top's direction of
motion in the lab frame, which we will refer to as the detection axis.
The most natural experimental observable for hadronic tops is then the angular
distribution of the direction of the bottom jet (or the less energetic
light quark jet) with respect to the top quark direction of motion in
the top rest frame,
\begin{equation}
\frac{1}{\Gamma} \frac{d\Gamma}{d \cos \theta'} = \onehalf \left( 1+ \langle \mcp_D \rangle \kappa \cos\theta'\right),
\end{equation}
The observed polarization is thus along the detection axis,
$\langle\mcp_D\rangle$.  For highly boosted tops, the natural polarization
axis is again the top direction of motion.  Notice that there is no 
interference between the two top spin states, as the observables 
we consider are insensitive to rotations around the top direction of motion. 

\begin{figure} 
\begin{center} 
\vspace{.25cm}
\includegraphics[width=8.0cm]{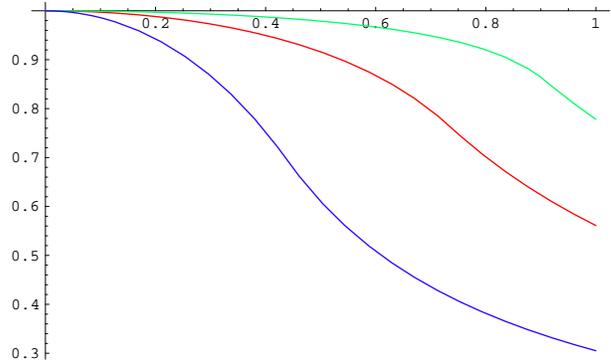} 
\end{center} 
%\vspace{-0.75cm} 
\begin{small}
\caption{Polarization reduction factor $\cos\omega$ for three
different sets of new particle masses, as a function of parent top
partner boost $\beta $.  The phase space integral over $d\cos\theta$
has been performed.  The red (central) curve is for a top partner with
mass $500\gev $ and boson partner $150\gev $.  The green
(upper) curve is for a top partner with mass $900\gev $ and
boson partner $300\gev $.  The blue (lower) curve is for a top partner
with mass $900\gev $ and boson partner $700\gev $.
\label{fig:omega}}
\end{small}
\end{figure}

In order to relate the observed polarization
to the polarization along the production axis $\langle\mcp_P\rangle $
computed in the previous section, we need to relate the two spin
bases.  The detected polarization $\langle \mcp_D\rangle$ is related
to the production polarization $\langle\mcp_P\rangle $ by
\begin{equation}
\langle \mcp_D\rangle=\langle \mcp_P\rangle\cos\omega,
\end{equation}
where $\cos\omega $ is the angle between the production and detection
axes in the top rest frame.  This angle is the Wigner angle determined
by the composition of boosts $\Lambda_{tL} $ from the top rest frame
to the lab frame, followed by $\Lambda_{LP} $ from the lab frame to
the parent rest frame, and finally $\Lambda_{Pt} $ from the parent
rest frame to the top rest frame,
\begin{equation}
\mcm (\Lambda_{tL})\mcm (\Lambda_{LP})\mcm (\Lambda_{Pt}) =\mc{R} (\omega),
\end{equation}
where $\mcm (\Lambda),\mc{R}(\theta) $ are the representation matrices
for the Lorentz transformations \cite{ms}.  From this composition of
boosts we find
\begin{equation}
\cos\omega =\frac{E_t\beta\cos\theta +|p_t|}{\sqrt{|p_t| ^ 2 (1+\beta ^ 2
\cos ^2\theta) +m_t ^ 2\beta ^ 2+2 E_t |p_t|\beta\cos\theta}},
\end{equation}
where $E_t, p_t $ are the energy and momentum of the top in the parent
rest frame, $ \theta $ is the angle of the top in the parent rest
frame, and $\beta $ is the boost between the parent rest frame and the
lab frame.  The energy and momentum of the top are fixed by the masses
of the particles in the decay.  The phase space integration over $\cos
\theta$ can be carried out simply as (after averaging over top partner and
vector boson partner spins) the matrix element is independent of $\cos
\theta$.  This leaves the boost $\beta $ between the parent rest frame
and the lab frame.  In the limit where the top
partner is produced near threshold with little velocity in the lab
frame, $\beta\to 0$ and $\cos\omega \to 1$, as there is little
difference between the production and detection spin axes.  The Wigner 
angle is also lessened when the top quark is highly boosted in the top partner
rest frame.  Thus, as expected, the maximal observable polarization 
signals are realized when the top partner is heavy and the vector boson 
partner is light.
We plot $\cos\omega$ as a function of $\beta $ in three different 
scenarios for the new particle masses in Figure~\ref{fig:omega} 
(we integrate over $\cos\theta$).
%%%%%%%%%%%%%%%%%%%%%%%%%%%%%%%%%%%%%%%%%%%%%%%%%%%%%%%%
%% Figure: Wigner angle as a function of boost
%%%%%%%%%%%%%%%%%%%%%%%%%%%%%%%%%%%%%%%%%%%%%%%%%%%%%%%%
%%%%%%%%%%%%%%%%%%%%%%%%%%%%%%%%%%%%%%%%%%%%%%%%%%%%%%%%
For boosts $\beta \simeq 0.6$, we find that the scenarios with a moderate to 
sizable mass splitting have $\cos \omega \sim 0.85$ or better (the red and green 
curves shown in the figure), while in the case of heavy, degenerate
standard model partners (the blue curve) the suppression is much larger, 
$\cos \omega \sim 0.5$.  For smaller boosts $\beta \simeq 0.25$ the situation
is much better, with a suppression $\cos\omega\sim 0.9$ even for the
heavy, degenerate scenario.  In principle
for a given model the distribution of boosts $\mcd (\beta)$ can be
computed from the production dynamics and the PDFs, and the observable
polarization is then given by
\begin{equation}
\langle \mcp_D\rangle=\langle \mcp_P\rangle \frac{1}{2}\int d\cos\theta \,
d\beta \,\mcd (\beta) \, \cos\omega (\theta,\beta),
\end{equation}
allowing detailed quantitative predictions.  In practice, for sufficiently 
heavy top partners (and a sufficiently large mass splitting between the top 
partner and its daughter particles), the detailed boost distribution may 
not be necessary.

%%%%%%%%%%%%%%%%%%%%%%%%%%%%%%%%%%%%%%%%%%%%%%%%%%%%%%%%
\section{Summary}
\label{sec:theend}
%%%%%%%%%%%%%%%%%%%%%%%%%%%%%%%%%%%%%%%%%%%%%%%%%%%%%%%%

Top quarks produced in association with new physics can serve as an
important window into the chiral structure of physics beyond the Standard Model.
We have computed here the expected top polarization arising from a well-motivated
class of events, namely on-shell cascade decays of a top partner into a top quark
and a vector boson partner.  Observation of this signal constrains both the
masses of the particles in the cascade and the chirality of the couplings.

We have focused on variables which can be used in events which are not 
fully reconstructed.  In the happy circumstance that the LHC reveals a rich spectrum 
of new particles with complicated cascade decays, this strategy is likely to prove 
the cleanest way to assemble a large sample of inclusive top + new physics events, 
particularly if the new physics includes a stable invisible particle.  We have
discussed several angular distributions relevant for hadronic top quarks, and 
for collimated tops have constructed variables for both hadronic and leptonic
decay modes and plotted their distributions.  

%An interesting avenue for future 
%work is the construction of polarization-sensitive observables for fully collimated 
%hadronic top quarks, when the $b$-quark (or $W$) cannot be separately resolved 
%within the top jet.

%%%%%%%%%%%%%%%%%%%%%%%%%%%%%%%%%%%%%%%%%%%%%%%%%%%%%%%%
\begin{acknowledgments}
The author would like to thank Scott Thomas, Matt Strassler, and Michael 
Graesser for useful discussions.  This work was supported in part
by DOE grant DE-FG02-96ER40959.
\end{acknowledgments}
%%%%%%%%%%%%%%%%%%%%%%%%%%%%%%%%%%%%%%%%%%%%%%%%%%%%%%%%

% ---------------------------------------------------------------------

\end{document}